\documentclass[journal]{IEEEtran}

\usepackage{amssymb, amsmath, amsthm}
\usepackage{graphicx, subfigure}
\usepackage{url, cite, verbatim}
\usepackage{booktabs}
\usepackage{xcolor}
\usepackage[ruled,vlined,linesnumbered]{algorithm2e}

\ifCLASSINFOpdf
\else
\fi
\newcommand{\ls}[1]  
   {\dimen0=\fontdimen6\the=#1\dimen0
    \advance\lineskip.5\fontdimen5\the\lineskip-\dimen0
    \lineskiplimit=.9\lineskip
    \baselineskip=\lineskip
    \advance\baselineskip\dimen0
    \normallineskip\lineskip
    \normallineskiplimit\lineskiplimit
    \normalbaselineskip\baselineskip
    \ignorespaces
   }

\begin{document}

\title{CSI-based Fingerprinting for Indoor Localization: A Deep Learning Approach}
 

\author{Xuyu~Wang,~\IEEEmembership{Student~Member,~IEEE,}
        Lingjun~Gao,~\IEEEmembership{Student~Member,~IEEE,}
        Shiwen~Mao,~\IEEEmembership{Senior~Member,~IEEE,}
        and~Santosh~Pandey
\thanks{Manuscript received Oct. 7, 2015; revised Feb. 1, 2016; accepted Mar. 18, 2016. 
This work is supported in part by the US National Science Foundation under Grant CNS-1247955 and the Wireless Engineering Research and Education Center (WEREC) at Auburn University. 
This works was presented in part at IEEE WCNC 2015, New Orleans, LA, Mar. 2015.}
\thanks{X. Wang, L. Gao, and S. Mao are with the Department of Electrical and Computer Engineering, Auburn University, Auburn, AL 36849-5201 USA. 
E-mail: \{xzw0029, lzg0014\}@auburn.edu, smao@ieee.org}%
\thanks{S. Pandey is with Cisco Systems, Inc., 170 West Tasman Dr., San Jose, CA 95134, USA. 
E-mail: sanpande@cisco.com}
\thanks{Copyright\copyright 2016 IEEE. Personal use of this material is permitted. However, permission to use this material for any other purposes must be obtained from the IEEE by sending a request to pubs-permissions@ieee.org.}%
}

\markboth{IEEE Transactions on Vehicular Technology, VOL. XX, NO. XX, MONTH YEAR}%
{Wang \MakeLowercase{\textit{et al.}}: CSI based Fingerprinting for Indoor Localization}

\maketitle

\begin{abstract}
With the fast growing demand of location-based services in indoor environments, indoor positioning based on fingerprinting has attracted a lot of interest due to its high accuracy. In this paper, we present a novel deep learning based indoor fingerprinting system using Channel State Information (CSI), which is termed DeepFi. Based on three hypotheses on CSI, the DeepFi system architecture includes an off-line training phase and an on-line localization phase. In the off-line training phase, deep learning is utilized to train all the weights of a deep network as fingerprints. Moreover, a greedy learning algorithm is used to train the weights layer-by-layer to reduce complexity. In the on-line localization phase, we use a probabilistic method based on the radial basis function to obtain the estimated location. Experimental results are presented to confirm that DeepFi can effectively reduce location error compared with three existing methods in two representative indoor environments.
\end{abstract}





\section{Introduction} \label{sec:intro}

With the proliferation of mobile devices, indoor localization has become an increasingly important problem. Unlike outdoor localization, such as the Global Positioning System (GPS),  that has line-of-sight (LOS) transmission paths, indoor localization faces a challenging radio propagation environment, including multipath effect, shadowing, fading and delay distortion~\cite{KM,XWang14}. In addition to the high accuracy requirement, an indoor positioning system should also have a low complexity and short online process time for mobile devices. To this end, fingerprinting-based indoor localization becomes an effective method to satisfy these requirements, where an enormous amount of measurements are essential to build a database to facilitate real-time position estimation.  

Fingerprinting based localization usually consists of two basic phases: (i) the off-line phase, which is also called the training phase, and (ii) the on-line phase, which is also called the test phase~\cite{Radar}. The training phase is for database construction, when survey data related to the position marks is collected and pre-processed. 
In the off-line training stage, machine learning methods can be used to train fingerprints instead of storing all the received signal strength (RSS) data. Such machine learning methods not only reduce the computational complexity, but also obtain the core features in the RSS for better localization performance. $K$-nearest-neighbor (KNN), neural networks, and support vector machine, as popular machine learning methods, have been applied for fingerprinting based indoor localization. KNN uses the weighted average of $K$ nearest locations to determine an unknown location with the inverse of the Euclidean distance between the observed RSS measurement and its $K$ nearest training samples as weights~\cite{KM}. A limitation of KNN is that it needs to store all the RSS training values. Neural networks utilizes the back-propagation algorithm to train weights, but it considers one hidden layer to avoid error propagation in the training phase and needs labeled data as a supervised learning~\cite{Cooperative}. Support vector machine uses kernel functions to solve the randomness and incompleteness of the RSS values, but has high computing complexity~\cite{Location}. 
In the on-line phase, a mobile device records real time data and tests it using the database. The test output is then used to estimate the position of the mobile device, by searching each training point to find the most closely matched one as the target location. Besides such nearest estimation method, an alternative matching algorithm is to identify several close points each with a maximum likelihood probability, and to calculate the estimated position as the weighted average of the candidate positions.

Many existing indoor localization systems use RSS as fingerprints due to its simplicity and low hardware requirements. For example, the Horus system uses a probabilistic method for location estimation with RSS data~\cite{Horus}. Such RSS based methods have two disadvantages. First, RSS values usually have a high variability over time for a fixed location, due to the multipath effects in indoor environments. Such high variability can introduce large location error even for a stationary device. Second, RSS values are coarse information, which does not exploit the many subcarriers in an orthogonal frequency-division multiplexing (OFDM) for richer multipath information. It is now possible to obtain channel state information (CSI) from some WiFi network interface cards (NIC), which can be used as fingerprints to improve the performance of indoor localization~\cite{CSI-Based, Predictable, Wang15a, WangGC15, WangIOT16}. For instance, the FIFS scheme uses the weighted average CSI values over multiple antennas~\cite{FIFS}. 
In addition, the PinLoc system also exploits CSI information, while considering $1\times1$ m$^2$ spots for training data~\cite{You}.

In this paper, we propose a deep learning based fingerprinting scheme to mitigate the several limitations of existing machine learning based methods. The deep learning based scheme can fully explore the feature of wireless channel data and obtain the optimal weights as fingerprints. It also incorporates a greedy learning algorithm to reduce computational complexity, which has been successfully applied in image processing and voice recognition~\cite{IN}. The proposed scheme is based on CSI to obtain more fine-grained information about the wireless channel than RSS based schemes, such as the amplitude and phase of each subcarrier from each antenna for each received packet. The proposed scheme is also different from the existing CSI based schemes, in that it incorporates 90 magnitudes of CSI values collected from the three antennas of the Intel's IWL 5300 NIC to train the weights of a deep network with deep learning. 

In particular, we present DeepFi, a deep learning based indoor fingerprinting scheme using CSI. We first introduce the background of CSI and present three hypotheses on CSI. We then present the DeepFi system architecture, which includes an off-line training phase and an on-line localization phase. In the training phase, CSI information for all the subcarriers from three antennas are collected from accessing the device driver and are analyzed with a deep network with four hidden layers. We propose to use the weights in the deep network to represent fingerprints, and to incorporate a greedy learning algorithm  using a stack of RBMs to train the deep network in a layer-by-layer manner to reduce the training complexity. The greedy algorithm first estimates the parameters of the first layer RBM to model the input data. Then the parameters of the first layer are frozen, and we obtain the samples from the conditional probability to train the second layer RBM and so forth. Finally, we can obtain the parameters of the fourth layer RBM with the above greedy learning algorithm. Moreover, for each layer RBM model, we use the contrastive divergence with 1 step iteration (CD-1) method to update weights, which has lower time complexity than other schemes such as Markov chain Monte Carlo (MCMC)~\cite{GD}. 
In the on-line localization phase, a probabilistic data fusion method based on radial basis function is developed for online location estimation using multiply packets. 
To reduce the computational complexity for online localization, packets are divided into several batches, each of which contains the same number of packets. Because packets are processed in parallel in batches, we can significantly shorten the processing time when dealing with a large amount of packets.


The proposed DeepFi scheme is validated with extensive experiments in two representative indoor environments, i.e., a living room environment and a computer laboratory environment. DeepFi is shown to outperform several existing RSSI and CSI based schemes in both experiments. We also examine the effect of different DeepFi parameters on localization accuracy and execution time, such as using different number of antennas, using different number of test packets, and different number of packets per batch.  
Finally, we investigate the effect of different propagation environments on the DeepFi performance, 
such as replaced obstacles, human mobility, and the training grid size in our experimental study. 
Our experimental results confirm that DeepFi can perform well in these scenarios. 

The remainder of this paper is organized as follows. The background and hypotheses are presented in Section~\ref{sec:sysMod}. The DeepFi system is presented in Section~\ref{sec:ad} and evaluated in in Section~\ref{sec:sml}. We review related work in Section~\ref{sec:RW} and Section~\ref{sec:conC} concludes this paper.

\section{Background and Hypotheses}\label{sec:sysMod}


\subsection{Channel State Information}

Thanks to the NICs, such as Intel's IWL 5300, it is now easier to conduct channel state measurements than in the past when one has to detect hardware records for physical layer (PHY) information. 
Now CSI can be retrieved from a laptop by accessing the 
device drive. 
CSI records the channel variation experienced during propagation. 
Transmitted from a source, a wireless signal may experience abundant impairments caused by, e.g., the multipath effect, fading, shadowing, and delay distortion. Without CSI, it is hard to reveal the channel characteristics with the signal power only.

Let $\vec{X}$ and $\vec{Y}$ denote the transmitted and received signal vectors. We have
\begin{eqnarray} \label{eq:CSI1}
  \vec{Y}=\mbox{CSI} \cdot \vec{X}+\vec{N}, 
\end{eqnarray}
where vector $\vec{N}$ is the additive white Gaussian noise and $\mbox{CSI}$ represents the channel's frequency response, which can be estimated from $\vec{X}$ and $\vec{Y}$. 

The WiFi channel at the 2.4 GHz band can be considered as a narrowband flat fading channel for OFDM system. The Intel WiFi Link 5300 NIC implements an OFDM system with 56 subcarriers, 30 out of which can be read for CSI information via the device driver. The channel frequency response $\mbox{CSI}_i$ of subcarrier $i$ is a complex value, which is defined by
\begin{eqnarray} \label{eq:CSI 3}
  \mbox{CSI}_i=|\mbox{CSI}_i|\exp{\{j {\angle{\mbox{CSI}_i}}\}}. 
\end{eqnarray}
where $|\mbox{CSI}_i|$ and $\angle{\mbox{CSI}_i}$ are the amplitude response and the phase response of subcarrier $i$, respectively. 
In this paper, the proposed DeepFi framework is based on these 30 subcarriers (or, CSI values), 
which can reveal much richer channel properties than RSSI.

\subsection{Hypotheses}

We next present three hypotheses about the CSI data, which are validated with the statistical results through our measurement study.


\subsubsection{Hypothesis 1} {\em CSI amplitude values exhibit great stability for continuously received packets at a fixed location, compared with RSS values.}

\begin{figure} [!t]
	\centering
\includegraphics[width=3.2in, height=2.1in]{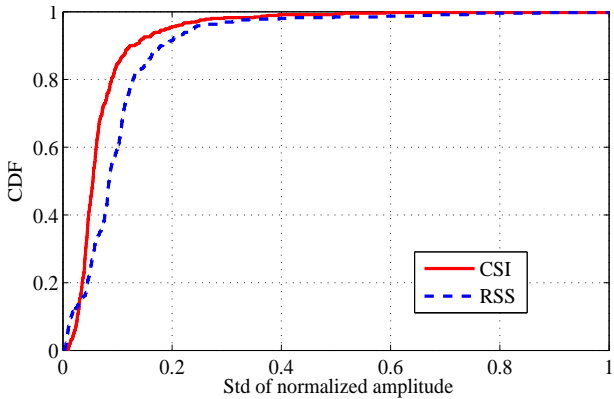}
\caption{CDF of the standard deviations of CSI and RSS amplitudes for 150 sampled locations.}
\label{fig:STD}
\end{figure}

CSI amplitude values reflect channel properties in the frequency domain and exhibit great stability over time for a given location.
Fig.~\ref{fig:STD} plots the cumulative distribution function (CDF) of the standard deviations of normalized CSI and RSS amplitudes for 150 sampled locations. At each location, CSI and RSS values are measured from 50 received packets with the three antennas of Intel WiFi Link 5300 NIC. It can be seen that for CSI amplitude values, 90\% of the standard deviations are blow 10\% of the average value. For RSS values, however, 60\% of the standard deviations are blow 10\% of the average value. Therefore, CSI is much more stable than RSS. Our measurements last a long period of time including both office hours and quiet hours. No obvious difference in the stability of CSI for the same location is observed at different times. On the contrary, RSS values exhibit large variations even at the same position. Therefore, CSI amplitude values are leveraged as the feature of deep learning in the DeepFi system. 


\subsubsection{Hypothesis 2}  {\em Because of the multipath effect and channel fading indoor environment, the number of clusters of CSI values over subcarriers varies at different locations.}

\begin{figure} [!t]
	\centering
\includegraphics[width=3.2in]{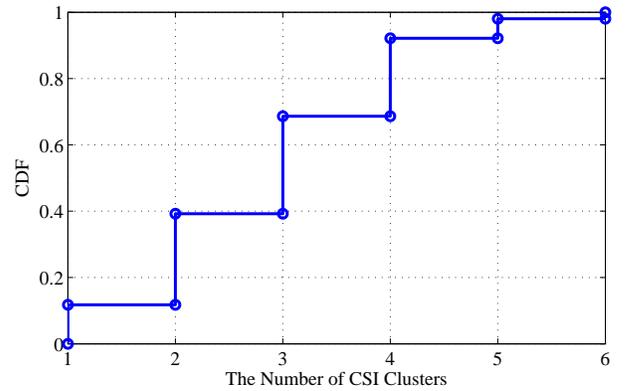}
\caption{CDF of the number of clusters of CSI amplitude values at 50 different locations.}
\label{fig:cluster}
\end{figure}

\begin{figure} [!t]
	\centering
\includegraphics[width=3.2in]{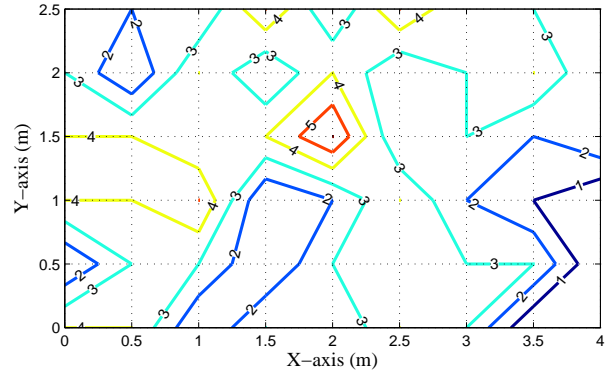}
\caption{2D contour of the number of clusters of CSI amplitude values at 50 different locations.}
\label{fig:contour}
\end{figure}

CSI amplitude values reflect channel frequency responses with abundant multipath components and channel fading. Our study of channel frequency responses shows that there are several dominant clusters for CSI amplitude values, where each cluster consists of a subset of subcarriers with similar CSI amplitudes values. To find the feature of clusters of CSI amplitudes values, we draw the CDF and the two-dimensional (2D) contour of the number of clusters for CSI amplitude values for 50 different locations in the living room environment in Figs.~\ref{fig:cluster} and~\ref{fig:contour}, respectively. For every location, CSI values are measured from 50 received packets with the three antennas of Intel WiFi Link 5300 NIC. From Figs.~\ref{fig:cluster} and~\ref{fig:contour}, we can see that the number of clusters of CSI amplitude values varies at 50 different locations. Moreover, at most of locations, CSI amplitude values form two or three clusters. Some locations have one cluster because of less reflection and diffusion. Some other locations with few five or six clusters may suffer from the severe multipath effect.

To detect all possible numbers of clusters, we measure CSI amplitude values from received packets for a long period of time at each location, which can be used for training weights in deep network. In addition, more packet transmissions will be helpful to reveal the comprehensive properties at each location. In our experiments, we consider 500 and 1000 packets for training in the living room environment and the computer laboratory environment, respectively, more than the 60 packets used in FIFS. 

\subsubsection{Hypothesis 3} {\em The three antennas of the Intel WiFi Link 5300 NIC have different CSI features, which can be exploited to improve the diversity of training and test samples.} 

\begin{figure} [!t]
	\centering
		\includegraphics[width=3.2in]{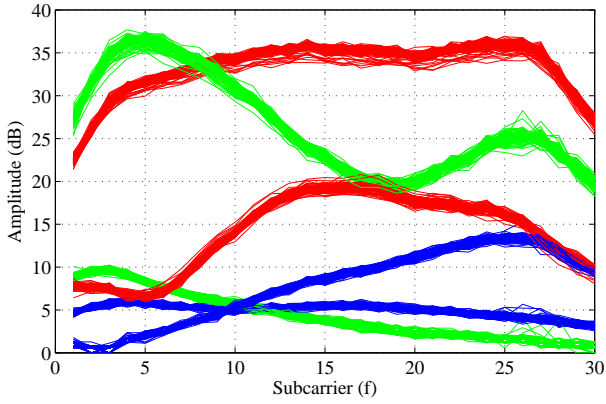}
\caption{Amplitudes of channel frequency response measured at the three antennas of the Intel WiFi Link 5300 NIC (each is plotted in a different color) for 50 received packets.}
\label{fig:subm}
\end{figure}

Intel WiFi Link 5300 is equipped with three antennas. We find that the channel frequency responses of the three antennas are highly different, even for the same packet reception. In Fig.~\ref{fig:subm}, amplitudes of channel frequency response from the three antennas 
exhibit different properties. 
In FIFS, CSI amplitude values from the three antennas are simply accumulated to produce an average value. In contrast, DeepFi aims to utilize their variability
to enhance the training and test process in deep learning. 
The 30 subcarriers can be treated as 30 nodes and used as input data of visible variability for deep learning. 
With the three antennas, 
there are 90 nodes 
that can be used as input data for deep learning. 
The greatly increased number of nodes for input data can improve the diversity of training and test samples, leading to better performance of localization if reasonable parameters are chosen.

\begin{figure} [!t]
	\centering
\includegraphics[width=3.0in]{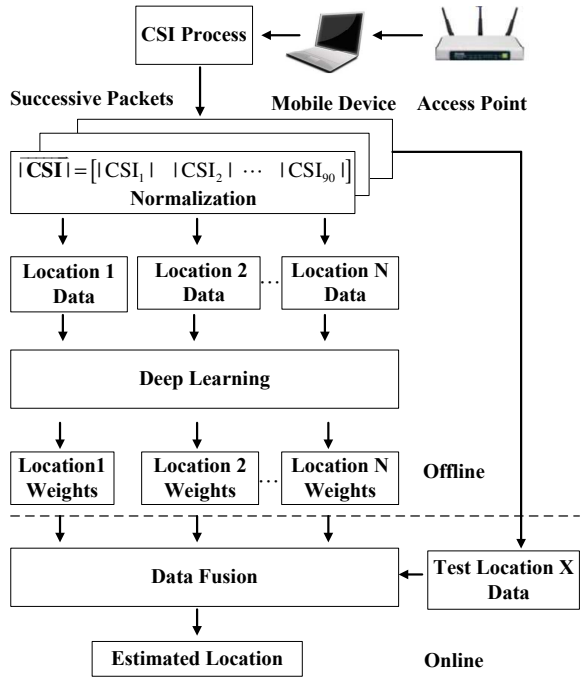}
\caption{The DeepFi architecture.}
\label{fig:DP}
\end{figure}

\section{The DeepFi System}\label{sec:ad}


\subsection{System Architecture}

Fig.~\ref{fig:DP} shows the system architecture of DeepFi, 
which only requires one access point and one mobile device equipped with an Intel WiFi link 5300 NIC. 
At the mobile device, raw CSI values can be read from the modified chipset firmware for received packets. 
The Intel WiFi link 5300 NIC has three antennas, each of which can collect CSI data from 30 different subcarriers. We can thus obtain 90 raw CSI measurements for each packet reception. Unlike FIFS that averages over multiple antennas to reduce the received noise, our system uses all CSI values from the three antennas for indoor fingerprint to exploit diversity of the channel. 
Since it is hard to use the phases of CSI for localization, we only consider 
the amplitude responses 
for fingerprinting in this paper. 
On the other hand, since the input values should be limited in the range (0, 1) for effective deep learning, we normalize the amplitudes of the 90 CSI values 
for both the offline and online phases.  

In the offline training phase, DeepFi generates feature-based fingerprints, which are highly different from traditional methods that directly store CSI values. Feature-based fingerprints utilize a large number of weights obtained by deep learning for different locations, which effectively describe the characteristics of CSI for each location and reduce noise. Meanwhile these weights can indirectly extract the feature of clusters hidden in CSI values. The feature-based fingerprints server can store the weights for different training locations. In the online localization phase, the mobile device can estimate its position with a data fusion approach, 
which will be described in Section~\ref{subsec:DF}. 

\subsection{Weight Training with Deep Learning \label{subsec:DL}}

\begin{figure} [!t]
	\centering
\includegraphics[width=3.0in]{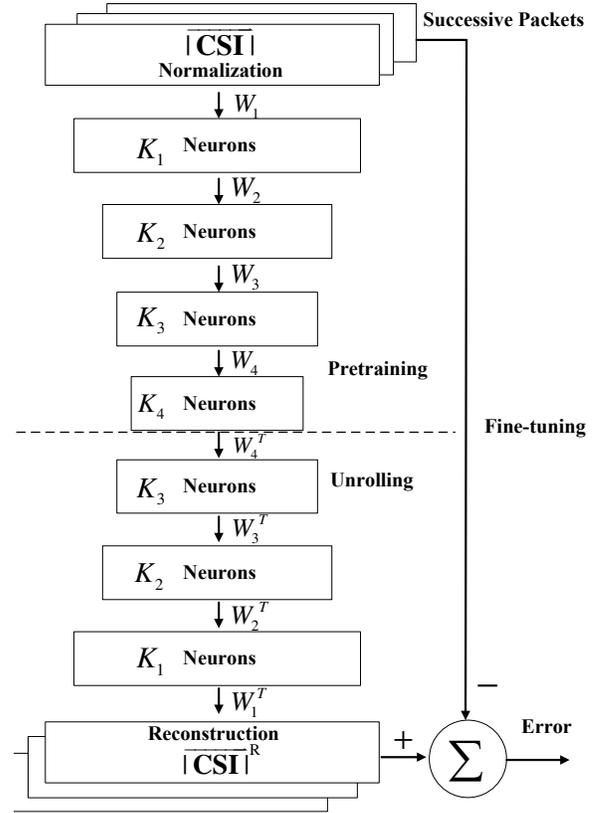}
\caption{Weight training with deep learning.}
\label{fig:train}
\end{figure}

Fig.~\ref{fig:train} illustrates how to train weights based on deep learning. There are three stages in the procedure, including pretraining, unrolling, and fine-tuning~\cite{Reducing}. 
A deep network with four hidden layers is adopted, where every hidden layer consists of a different number of neurons. In order to reduce the dimension of CSI data, we assume that the number of neurons in a higher hidden layer is more than that in a lower hidden layer. Let $K_{1}, K_{2}, K_{3}$ and $K_{4}$ denote the number of neurons in the first, second, third, and fourth hidden layer, respectively. 
It follows that $K_{1}>K_{2}>K_{3}>K_{4}$. 

In addition, we propose a new approach to represent fingerprints, i.e., using the weights between connected layers. Define $W_{1}, W_{2}, W_{3}$ and $W_{4}$ as the weights between the normalized magnitudes of CSI values and the first hidden layer, the first and second hidden layer, the second and third hidden layer, and the third and fourth hidden layer, respectively. The key idea is that after training the weights in the deep network, we can store them as fingerprints to facilitate localization in the on-line test stage. Moreover, we define $h_{i}$ as the hidden variable at layer $i$, for $i=1,2,3,4$, respectively, and let $v$ denote the input data, i.e., the normalized CSI magnitudes. 

We represent the deep network with four hidden layers with a probabilistic generative model, which 
can be written as
\begin{align} \label{eq:joint distribution1}
  &\; \Pr(v,h^{1},h^{2},h^{3},h^{4}) \nonumber \\
	= &\; \Pr(v|h^{1})\Pr(h^{1}|h^{2})\Pr(h^{2}|h^{3})\Pr(h^{3},h^{4}). 
\end{align}
Since the nodes in the deep network are mutually independent, $\Pr(v|h^{1})$, $\Pr(h^{1}|h^{2})$, and $\Pr(h^{2}|h^{3})$ can be represented by 
\begin{align} \label{eq:conditional distribution1}
  \left\{   \begin{array}{l}
	          \Pr(v|h^{1})=\prod_{i=1}^{90}{\Pr(v_i|h^1)} \\
	          \Pr(h^{1}|h^{2})=\prod_{i=1}^{K_1}{\Pr(h_{i}^1|h^2)} \\  
            \Pr(h^{2}|h^{3})=\prod_{i=1}^{K_2}{\Pr(h_{i}^2|h^3)}. 
						     \end{array} \right. 
\end{align}
In~(\ref{eq:conditional distribution1}), $\Pr(v_i|h^1)$, $\Pr(h_{i}^1|h^2)$, and $\Pr(h_{i}^2|h^3)$ are described by the sigmoid belief network in the deep network, as
\begin{align} \label{eq:sigmoid1}
  \left\{   \begin{array}{l}
			\Pr(v_i|h^1)= 1/\left(1+\exp{(-b_i^0-\sum_{j=1}^{K_1}{W_1^{i,j}h_j^1)}}\right)\\
			\Pr(h_{i}^1|h^2)= 1/\left(1+\exp{(-b_i^1-\sum_{j=1}^{K_2}{W_2^{i,j}h_j^2)}}\right)\\  
			\Pr(h_{i}^2|h^3)= 1/\left(1+\exp{(-b_i^2-\sum_{j=1}^{K_3}{W_3^{i,j}h_j^3)}}\right) 
					\end{array} \right. 
\end{align}
where $b_i^0$, $b_i^1$ and $b_i^2$ are the biases for unit $i$ of input data $v$, unit $i$ of layer 1, and unit $i$ of layer 2, respectively. 

The joint distribution $\Pr(h^{3},h^{4})$ can be expressed as an Restricted Bolzmann Machine (RBM)~\cite{GD} with a bipartite undirected graphical model~\cite{GD}, which is given by
\begin{eqnarray} \label{eq:joint distribution2}
  \Pr(h^{3},h^{4})=\frac{1}{Z} \exp(-\mathbb{E}(h^{3},h^{4})), 
\end{eqnarray}
where 
  $Z=\sum_{h^{3}}\sum_{h^4}{\exp(-\mathbb{E}(h^{3},h^{4}))}$ and 
	$\mathbb{E}(h^{3},h^{4})=-\sum_{i=1}^{K_3}{b_i^3h_i^3}-\sum_{j=1}^{K_4}{b_j^4h_j^3}-\sum_{i=1}^{K_3}\sum_{j=1}^{K_4}{W_4^{i,j}h_{i}^3h_{j}^4}$. 
In fact, since it is difficult to find the joint distribution $\Pr(h^{3},h^{4})$, we use CD-1 method~\cite{GD} to approximate it, which is given by
\begin{eqnarray} \label{eq:conditional distribution2}
\left\{   \begin{array}{l}
	\Pr(h^{3}|h^{4})=\prod_{i=1}^{K_3}{\Pr(h_{i}^3|h^4)} \\  
  \Pr(h^{4}|h^{3})=\prod_{i=1}^{K_4}{\Pr(h_{i}^4|h^3)}, 
              \end{array} \right. 
\end{eqnarray}
where $\Pr(h_{i}^3|h^4)$, and $\Pr(h_{i}^4|h^3)$ are described by the sigmoid belief network, as
\begin{eqnarray} \label{eq:sigmoid2}
\left\{   \begin{array}{l}
	\Pr(h_{i}^3|h^4)= 1 / \left( 1+\exp{(-b_i^3-\sum_{j=1}^{K_4}{W_4^{i,j}h_j^4)}}\right) \\  
  \Pr(h_{i}^4|h^3)= 1 / \left( 1+\exp{(-b_i^4-\sum_{j=1}^{K_3}{W_4^{i,j}h_j^3)}} \right). 
	              \end{array} \right. 
\end{eqnarray}
Finally, the marginal distribution of input data for the deep belief network is given by
\begin{eqnarray} \label{eq:sigmoid3}  
  \Pr(v)=\sum_{h^1}\sum_{h^2}\sum_{h^3}\sum_{h^4}\Pr(v,h^1,h^2,h^3,h^4).
\end{eqnarray}

Due to the complex model structure with the large number of neurons and multiple hidden layers in the deep belief network, it is difficult to obtain the weights using the given input data with the maximum likelihood method. In DeepFi, we adopt a greedy learning algorithm using a stack of RBMs to train the deep network in a layer-by-layer manner~\cite{GD}. This greedy algorithm first estimates the parameters $\{b^0,b^1,W_1\}$ of the first layer RBM to model the input data. Then the parameters $\{b^0,W_1\}$ of the first layer are frozen, and we obtain the samples from the conditional probability $\Pr(h^1|v)$ to train the second layer RBM (i.e., to estimate the parameters $\{b^1,b^2,W_2\}$), and so forth. Finally, we can obtain the parameters $\{b^3,b^4,W_4\}$ of the fourth layer RBM with the above greedy learning algorithm. 

For the layer $i$ RBM model, we use the CD-1 method to update weights $W_i$. We first get $h^i$ based on the samples from the conditional probability $\Pr(h^i|h^{i-1})$, and then obtain $\hat{h}^{i-1}$ based on the samples from the conditional probability $\Pr(h^{i-1}|h^i)$. Finally we obtain $\hat{h}^i$ using the samples from the conditional probability $\Pr(h^i|\hat{h}^{i-1})$. Thus, we can update the parameters as follows. 
\begin{eqnarray} \label{eq:sigmoid4}
\left\{   \begin{array}{l}
	W_i=W_i+\alpha(h^{i-1}h^{i}-\hat{h}^{i-1}\hat{h}^i)\\ 
	b^i=b^i+\alpha(h^{i}-\hat{h}^i)\\ 
  b^{i-1}=b^{i-1}+\alpha(h^{i-1}-\hat{h}^{i-1}), 
		              \end{array} \right. 
\end{eqnarray}
where $\alpha$ is the step size. After the pretraining stage, we need to unroll the deep network to obtain the reconstructed data $\hat{v}$ using the input data with forward propagation. The error between the input data $v$ and the reconstructed data $\hat{v}$ can be used to adjust the weights at different layers with the back-propagation algorithm. This procedure is called fine-tuning. By minimizing the error, we can obtain the optimal weights to represent fingerprints, which are stored in a database for indoor localization in the on-line stage.

The pseudocode 
for weight learning with multiply packets is given in Algorithm~\ref{tab:offline}. We first collect $m$ packet receptions for each of the $N$ training locations, each of which has 90 CSI values,  
as input data. Let $v(t)$ be the input data from packet $t$. The output of the algorithm consists of $N$ groups of fingerpirnts, each of which has eight weight matrices. 
In fact, we need to train a deep network for each of the $N$ training locations. The training phase includes three steps: pretraining, unrolling and fine-tuning. For pretraining, the deep network with four hidden layers is trained with the greedy learning algorithm. The weight matrix and bias of every layer are initialized first, and are then iteratively updated with the CD-1 method for obtaining a near optimal weight, where $m$ packets are trained and iteratively become output as input of the next hidden layer (lines 4-21). 

Once weight training 
is completed, the input data will be unrolled to obtain the reconstructed data. First, we use the input data to compute $\Pr(h^i|h^{i-1})$ based on the sigmoid with input $h^{i-1}$ to obtain the coding output $h^4$, which is a reduced dimension data (lines 23-26). Then, by computing $\Pr(\hat{h}^{i-1}|\hat{h}^{i})$ based on the sigmoid with input $\hat{h}^{i}$, we can sample the reconstructed data $\hat{h}^{0}$, where the weights of the deep network are only transposed, thus reducing the time complexity of weight learning (lines 27-31). Once the reconstructed data $\hat{h}^{0}$ if obtained, the unsupervised learning method for the deep network becomes a supervised learning problem as in the fine-tuning phase. Thus, we compute the error between the input data $v=h^{0}$ and reconstructed data $\hat{h}^{0}$ to successively update the weight matrix with the standard back-propagation algorithm (lines 33-34).

\begin{algorithm} [!t]
   \textbf{Input:} $m$ packet receptions each with 90 CSI values for each of the $N$ training locations\;
	 \textbf{Output:} $N$ groups of fingerprints each consisting of eight weight matrices\;
\small
\SetAlgoLined
	 \For{$j=1:N$}{
	     // pretraining\;			
		 \For{$i=1:4$}{
					Initialize $W^i=0.1 \cdot randn$, $b^i=0$,  // $randn$ is the standard Gaussian distribution function\; 
					\For{$k=1:$ maxepoch}{
					  \For{$t=1:m$}{
					 $h^{0}=v(t)$\;
					 Compute $\Pr(h^i|h^{i-1})$ based on the sigmoid with input $h^{i-1}$\;
					 Sample $h^i$ from $\Pr(h^i|h^{i-1})$\;
					 Compute $\Pr(h^{i-1}|h^i)$ based on the sigmoid with input $h^{i}$\;
					 Sample $\hat{h}^{i-1}$ from $\Pr(h^{i-1}|h^i)$\;
					 Compute $\Pr(h^i|\hat{h}^{i-1})$ based on the sigmoid with input $\hat{h}^{i-1}$\;
					 Sample $\hat{h}^i$ from $\Pr(h^i|\hat{h}^{i-1})$\;
					 $W_i=W_i+\alpha(h^{i-1}h^{i}-\hat{h}^{i-1}\hat{h}^i)$\;
	         $b^i=b^i+\alpha(h^{i}-\hat{h}^i)$\;
           $b^{i-1}=b^{i-1}+\alpha(h^{i-1}-\hat{h}^{i-1})$\; 				
					 }
					}
			}
			//unrolling\;	
			\For{$i=1:4$}{
					 Compute $\Pr(h^i|h^{i-1})$ based on the sigmoid with input $h^{i-1}$\;
					 Sample $h^i$ from $\Pr(h^i|h^{i-1})$\;
				}
			Set $\hat{h}^{i}=h^{i}$\;
		  \For{$i=4:1$}{
					 Compute $\Pr(\hat{h}^{i-1}|\hat{h}^{i})$ based on the sigmoid with input $\hat{h}^{i}$\;
					 Sample $\hat{h}^{i-1}$ from $\Pr(\hat{h}^{i-1}|\hat{h}^{i})$\;
				}
			//fine-tuning\;	
			Obtain the error between input data $h^{0}$ and reconstructed data $\hat{h}^{0}$ \;
			Update the eight weights using the error with back-propagation\;
			}
    \caption{Training for Weight Learning}
\label{tab:offline}
\end{algorithm}

\subsection{Location Estimation based on Data Fusion \label{subsec:DF}}

After off-line training, we need to test it with positions that are different from those used in the training stage. Because the probabilistic methods have better performance 
than deterministic ones, we use the probability model based on Bayes' law, which is given by
\begin{eqnarray} \label{eq:conditional distribution3}
  \Pr(L_{i}|v)= \frac{\Pr(L_{i})\Pr(v|L_{i}) }{ \sum_{i=1}^{N}{\Pr(L_{i})\Pr(v|L_{i})}}. 
\end{eqnarray}
In~(\ref{eq:conditional distribution3}), $L_{i}$ is reference location $i$, $\Pr(L_{i}|v)$ is the posteriori probability, $\Pr(L_{i})$ is the prior probability that the mobile device is determined to be at reference location $i$, and $N$ is the number of reference locations. In addition, we assume that $\Pr(L_{i})$ is uniformly distributed in the set $\{1, 2, \cdots,N\}$, and thus $\Pr(L_{i})=1/N$. It follows that
\begin{eqnarray} \label{eq:conditional distribution4}
  \Pr(L_{i}|v)=\frac{\Pr(v|L_{i})\frac{1}{N} }{ \sum_{i=1}^{N}{\Pr(v|L_{i})\frac{1}{N}}} 
	 =\frac{\Pr(v|L_{i}) }{ \sum_{i=1}^{N}{\Pr(v|L_{i})}}. 
\end{eqnarray}

Based on the deep network model, we define $\Pr(v|L_{i})$ as the radial basis function (RBF)  in the form of a Gaussian function, which is formulated as
\begin{eqnarray} \label{eq:conditional distribution5}
  \Pr(v|L_{i})=\exp\left(-\frac{\left\|v-\hat{v}\right\| }{ \lambda\sigma}\right), 
\end{eqnarray}
where $v$ is the input data, $\hat{v}$ is the reconstructed input data, $\sigma$ is the variance of input data, $\lambda$ is the coefficient of variation (CV) of input data. In fact, we use multiple packets to estimate the location of a mobile device, thus improving the indoor localization accuracy. For $n$ packets, we need to compute the average value of RBF, which is given by
\begin{eqnarray} \label{eq:conditional distribution55}
  \Pr(v|L_{i})=\frac{1}{n} \sum_{i=1}^{n}{\exp\left(-\frac{\left\|v_i-\hat{v}_i\right\| }{ \lambda\sigma}\right)}. 
\end{eqnarray}

Finally, the position of the mobile device can be estimated as a weighted average of all the reference locations, as 
\begin{eqnarray} \label{eq:conditional distribution6}
  \hat{L}=\sum_{i=1}^{N}{\Pr(L_{i}|v)L_{i}}. 
\end{eqnarray}

The pseudocode for online location estimation with multiply packets is presented in Algorithm \ref{tab:online}. The input to the algorithm consists of $n$ packet receptions, each of which has 90 CSI values, and $N$ groups of fingerprints obtained in the off-line training phase, each of which has eight weight matrices for each known training locations. 
First, we compute the variance 
of the 90 CSI values from each packet. We also group the $n$ packets into $a$ batches, each with $b$ packets, for accelerating the matching algorithm 
(lines 3-4). To obtain the posterior probability for different locations, we need to compute the RBF as likelihood function based on the reconstructed CSI values and input CSI values, where the reconstructed CSI values are obtained by recursively unrolling the deep network using the input data with forward propagation. For batch $j$, the reconstructed CSI values $\hat{V}_j$ are obtained by iterating the input data $V_j$ based on the eight weight matrices (lines 10-12). Then the sum of the RBFs (i.e., the $d_j$'s) is obtained by 
summing over the 90 CSI values and the $b$ packets in each batch (line 13). In addition, the expected RBF is computed by averaging over all the $n$ packets (line 16). Then, we compute the the posteriori probability $Pr_i$ for every reference location, thus obtaining the estimated position of the mobile device as the weighted average of all the reference locations (lines 19-23).

\begin{algorithm} [!t]
   \textbf{Input:} $n$ packet receptions each with 90 CSI values, $N$ groups of fingerprints each with eight weight matrices and the known training location\;
	 \textbf{Output:} estimated location $\hat{L}$\;	
\small
\SetAlgoLined
	 Compute the variance of CSI values $\sigma$\;
	 Group the $n$ packets into $a$ batches, each with $b$ packets\;
	 \For{$i=1:N$}{
		 \For{$j=1:a$}{
		      //compute the reconstructed CSI $\hat{V}_j$ with $b$ packets\;
          $\hat{V}_j=V_j$\;
					//where $V_j$ is the matrix with 90 rows and $b$ columns\;
				  \For{$k=1:8$}{
					 $\hat{V}_j= 1 / (1+\exp(-\hat{V}_j \cdot W_k))$\;
					}	
					 $d_j=\sum_{m=1}^b\exp\left(-\frac{1}{\lambda\sigma}\sum_{t=1}^{90}\sqrt{(V_j^{tm}-\hat{V}_j^{tm})^2} \right)$\;
					//where $V_j^{tm}$ is the element at row $t$ and column $m$ in matrix $V_j$, $\hat{V}_j^{tm}$ is the element at row $t$ and column $m$ in matrix $\hat{V}_j$\;
			}
			$P_i=\frac{1}{n} \sum_{j=1}^a d_j$\;	    
		}
		  // Obtain the posterior probability for different locations\;
			\For{$i=1:N$}{
		    $Pr_i= P_i / \sum_{i=1}^N P_i$\;
			}
			// Compute the estimated location\;
		  $\hat{L}=\sum_{i=1}^N Pr_i L_i$ \;	
    \caption{Online Location Estimation}
\label{tab:online}
\end{algorithm}

\section{Experiment Validation}\label{sec:sml}

\subsection{Experiment Methodology}

Our experiment testbed is implemented with two major components, the access point, which is a TP Link router, and the mobile terminal, which is a Dell laptop equipped with the IWL 5300 NIC. At the mobile device, the IWL 5300 NIC receives wireless signals from the access point, and then stores raw CSI values in the firmware. 
In order to read CSI values from the NIC driver, we install the 32-bit Ubuntu Linux, version 10.04LTS of the Server Edition on a Dell laptop and modify the kernel of the wireless driver. In the new kernel, raw CSI values can be transferred to the 
laptop and can be conveniently read with a C program. 

At the access point, the TL router is in charge of continuously transmitting packets to the mobile device. Since the router needs to respond to a mobile device who requires localization service, we use Ping to generate the request and response process between the laptop and the router. Initially, the laptop Pings the router, and then the router returns a packet to the laptop. In our experiment, we design a Java program to implement continuous Pings at a rate of 20 times per second. There are two reasons to select this rate. 
First, if we run Ping at a lower rate, no enough packets will be available to estimate a mobile device position. Second, if too many Pings are sent, there may not be enough time for the laptop to process the received packets. Also, since we need to continuously estimate the device position, it may cause buffer overflow and packet loss. 
In addition, after the IWL 5300 NIC receives a packet, the raw CSI value will be recorded in the hardware in the form of CSI per packet reception. DeepFi can obtain 90 raw CSI values for each packet reception, which are all used for fingerprinting or for estimating the device position.

We experiment with DeepFi and examine both the training phase and the test phase. During the training phase, CSI values collected at each location are utilized to learn features, which are then stored as fingerprints. In the test phase, we need to use online data to match the closest spot with the similar feature stored in the training phase. In fact, a major challenge in the feature matching is how to distinguish each spot without overlap or fuzziness. Although CSI features vary for different propagation paths, two spots with a shorter distance and a similar propagation path may have a similar feature. We examine the similarity of CSI feature along with spot interval in Section~\ref{subsec:sizespot}, 
where more details are discussed. If the training spots we select are too sparse, it is possible to cause fuzziness in the test phase, resulting in low localization accuracy. For example, a measurement could hardly match any training spot with high similarity, as it in fact has strong similarity with many random spots. On the other hand, if we choose dense training spots, it will cost a lot of efforts on pre-training data collection. Based on our experiments, the distance between two spots is set to 50 cm, which can maintain the balance between localization accuracy and pre-process cost. 

Since DeepFi fully explores all CSI features to search for the most matched spot, each packet is able to fit its nearest training spot with high probability. Therefore, in our localization system, only one access point is utilized to implement DeepFi, which can achieve similar precision as other methods such as Horus and FIFS with two or more access points. Although DeepFi has high accuracy with a single access point, it needs more time and computation in the offline training phase in order to learn 
fine-grained features of the spots. Fortunately, the pre-training process will be performed in the offline phase, while the online test phase can estimate position quickly. 
We design a data collection algorithm with two parts. 
In the training phase, we continuously collect 500--1000 packets at each spot and the measurement will lasts for 1 min. When collecting packets in our experiment, the laptop remains static on the floor, while all the test spots are at the same height, 
which construct a 2D platform. Then all the packets collected at each spot are used in DeepFi to calculate the weights of the deep network, which are stored as a spot feature. In the test phase, since we match for the closest position with weights we have saved in the database, it is unnecessary to group a lot of packets for complex learning processing. We thus use 100 packets to estimate position, 
thus significantly reducing the operating complexity and cost. 

We verify the performance of DeepFi in various scenarios and compare the resulting location errors in different environments with several benchmark schemes. We find that in an open room where there are no 
obstacles around the center, the performance of indoor localization is better than that in a complex environment where there are fewer LOS paths. We present the experimental results from two typical indoor localization environments, as described in the following.

\subsubsection{Living Room in a House}

\begin{figure} [!t]
	\centering  
\includegraphics[width=3.2in]{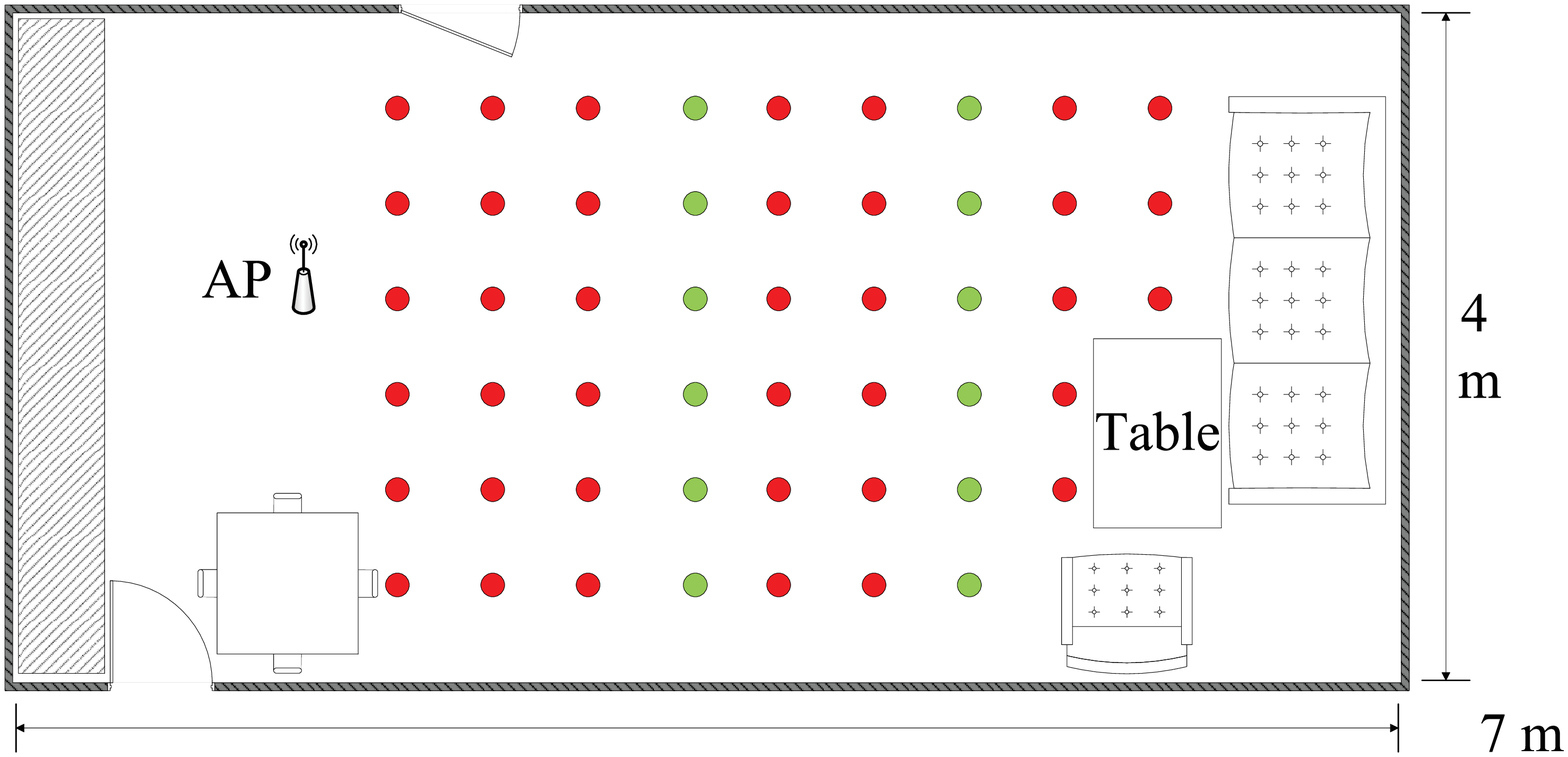}
\caption{Layout of the living room for training/test positions.}
\label{fig:home2002}
\end{figure}

The living room we choose is almost empty, so that most of the measured locations have LOS receptions. In this $4\times 7$ m$^2$ room, the access point was placed on the floor, and so do all the training and test points. As shown in Fig.~\ref{fig:home2002}, 50 positions are chosen uniformly scattered with half meter spacing in the room. Only one access point is utilized in our experiment, which is placed at one end (rather than the center) of the room to avoid isotropy. We arbitrarily choose 12 positions along two lines as test positions and use the remaining positions for training (in Fig.~\ref{fig:home2002}: the training positions are marked in red and the test positions are marked in green). For each position, we collect CSI data for nearly 500 packet receptions in 60 seconds. 
We choose a deep network with structure $K_1=300, K_2=150, K_3=100$, and $K_4=50$ 
for the living room environment. 

\subsubsection{Computer Laboratory}

\begin{figure} [!t]
	\centering
\includegraphics[width=3.2in]{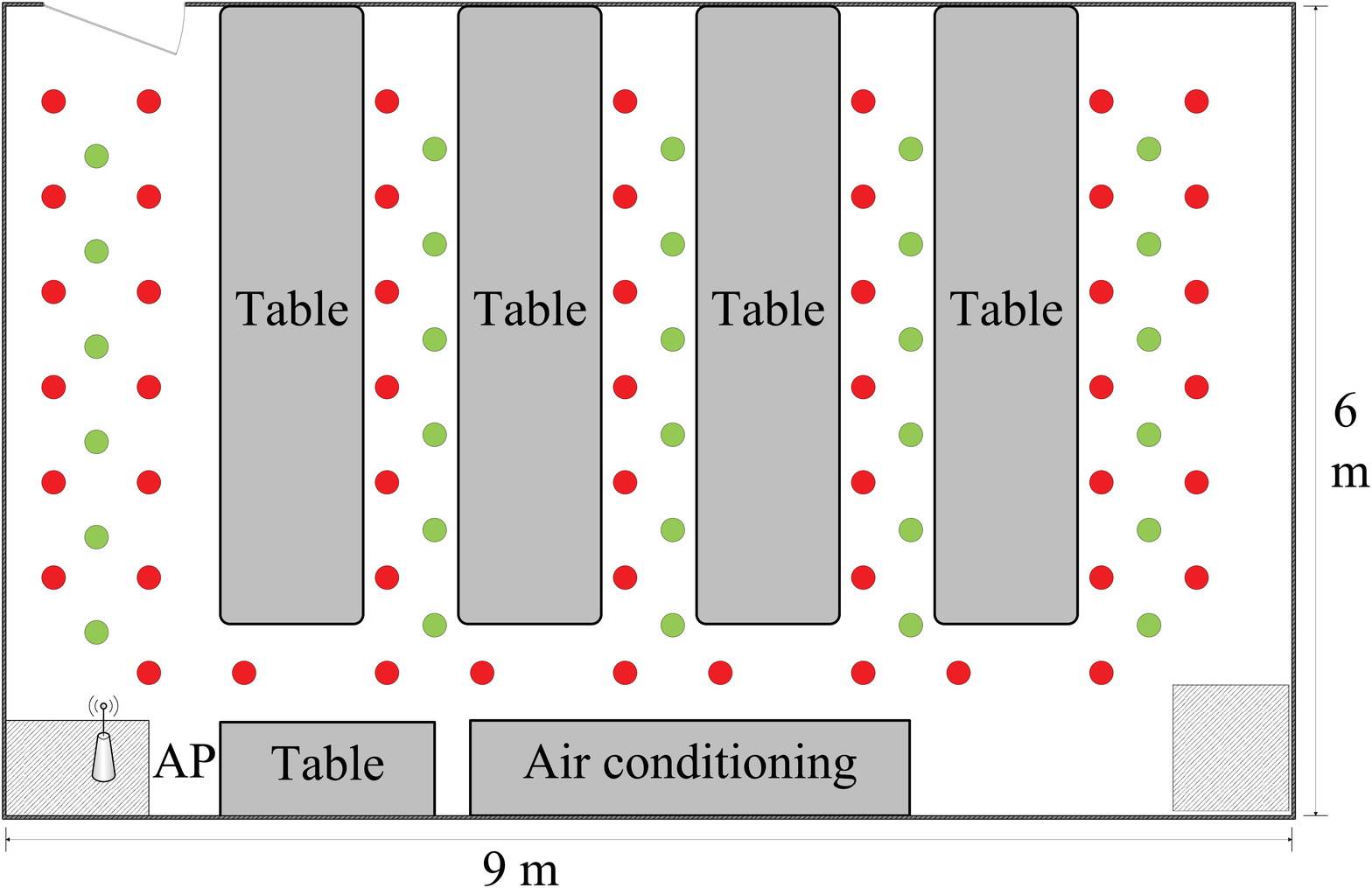}
\caption{Layout of the laboratory for training/test positions.}
\label{fig:lab2002}
\end{figure}

The other test scenario is a computer laboratory in Broun Hall in the campus of Auburn University. There are many tables and PCs crowded in the $6\times 9$ m$^2$ room, which block most of the LOS paths and form a complex radio propagation environment. In this laboratory, 50 training positions and 30 test positions are selected, as shown in Fig.~\ref{fig:lab2002}. The mobile device will also be put at these locations on the floor, with LOS paths blocked by the tables and computers. 
To obtain fine-grained characteristics of the subcarriers, CSI information from 1000 packet receptions are collected at each training position. 
We choose a deep network with structure $K_1=500, K_2=300, K_3=150$, and $K_4=50$
for the laboratory environment.

\subsubsection{Benchmarks and Performance Metric}

For comparison purpose, we implemented three existing methods, including FIFS~\cite{FIFS}, Horus~\cite{Horus}, and Maximum Likelihood (ML)~\cite{Statistical}. FIFS and Horus are introduced in Section~\ref{sec:intro}. In ML, the maximum likelihood probability is used for location estimation with RSS, where only one candidate location is used for the estimation result.
For a fair comparison, these schemes use the same measured dataset as DeepFi to estimate the location of the mobile device.   

The performance metric for the comparison of localization algorithms is the mean sum error $\mathcal{E}$. Assume the estimated location of an unknown user $i$ is $(\hat{x}_i, \hat{y}_i)$ and the actual position of the user is $(x_i, y_i)$. 
For $K$ locations, the mean sum error 
is computed as $\mathcal{E} = \frac{1}{K} \sum_{i=1}^K \sqrt{(\hat{x}_i-x_i)^2+(\hat{y}_i-y_i)^2}$. 

%

\subsection{Localization Performance}

We first evaluate the performance of DeepFi
under the two representative scenarios. The mean and standard deviation of the location errors are presented in Table~\ref{tb:error1}. 
In the living room experiment, the mean distance error is about 0.95 meter for DeepFi with a single access point. In the computer laboratory scenario, where there exists abundant multipath and shadowing effect, the mean error is about 1.8 meters across 30 test points. 
DeepFi outperforms FIFS in both scenarios; the latter has a mean error of 1.2 meters in the living room scenario and 2.3 meters in the laboratory scenario. DeepFi achieves a 20\% improvement over FIFS, 
by exploiting the fine-grained properties of CSI subcarriers from the three antennas. 
Both CSI fingerprinting schemes, i.e., DeepFi and FIFS, outperform the two RSSI-based fingerprinting schemes, i.e., Horus and ML. The latter two 
have errors of 2.6 and 2.8 meters, respectively, in the laboratory experiment. 

\begin{table} [!t]
\centering
\caption{Mean errors for the Living Room and and Laboratory Experiments}
\begin{tabular}{l|l|l|l|l}
\toprule
     &  \multicolumn{2}{|l|}{Living Room} & \multicolumn{2}{|l}{Laboratory} \\
\midrule
{\em Method} & {\em Mean error} & {\em Std. dev.}  & {\em Mean error} & {\em Std. dev.} \\
                               & {\em (m)}             & {\em (m)}           & {\em (m)}              & {\em (m)} \\
\midrule
DeepFi & $0.9425 $ & $0.5630$ & $1.8081 $ & $1.3432$ \\
FIFS  & $1.2436 $ & $0.5705$ & $2.3304 $ & $1.0219$ \\
Horus & $1.5449 $ & $0.7024$ & $2.5996 $ & $1.4573$ \\
ML & $2.1615 $ & $1.0416$ & $2.8478 $ & $1.5545$ \\
\bottomrule
\end{tabular}
\label{tb:error1}
\end{table}

\begin{figure} [!t]
	\centering
		\includegraphics[width=3.2in]{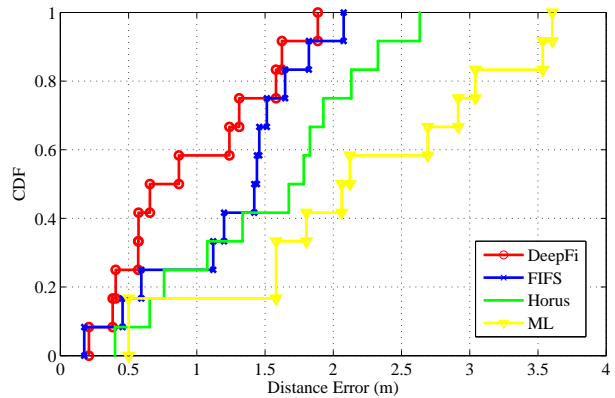}
\caption{CDF of localization errors in the living room experiment.}
\label{fig:home}
\end{figure}

Fig.~\ref{fig:home} presents the CDF of distance errors with the four methods in the living room experiment. 
With DeepFi, about 60\% of the test points have an error under 1 meter, 
while FIFS ensures that 
about 25\% of the test points have an error under 1 meter. In addition, most of the test points have distance errors less than 1.5 meters in FIFS, which is similar to DeepFi. On the other hand, both RSSI methods, 
i.e., Horus and ML, do not perform as well as the CSI-based schemes. There are only 80\% of the points have an error under 2 meters.

\begin{figure} [!t]
	\centering
		\includegraphics[width=3.2in]{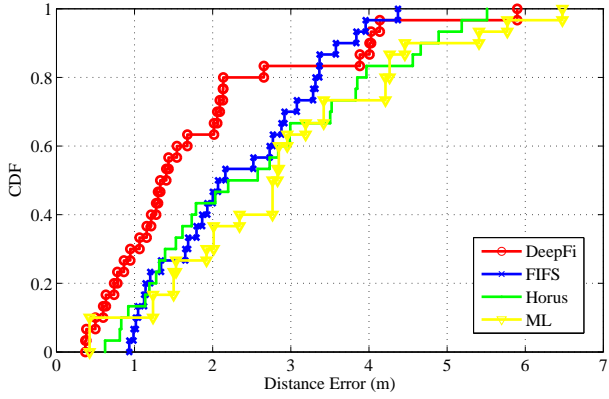}
\caption{CDF of localization errors in the laboratory experiment.}
\label{fig:lab}
\end{figure}

Fig.~\ref{fig:lab} plots the CDF of distance errors in the laboratory experiment. 
In this more complex propagation environment, DeepFi can achieve a 1.7 meters distance error for over 60\% of the test points, which is the most accurate one among the four schemes. Because the tables obstruct most LOS paths and magnify the multipath effect, 
the correlation between signal strength and propagation distance is weak in this scenario. The methods based on propagation properties, i.e., FIFS, Horus, and ML all have degraded performance than in the living room scenario. 
In Fig.~\ref{fig:lab}, it is noticed that 70\% of the test points have a 3 meters distance error with FIFS and Horus.
Unlike FIFS, DeepFi exploits various CSI subcarriers. It achieves higher accuracy even with just a single access point. It performs well in this NLOS environment. 

\subsection{Effect of Different System Parameters}

\subsubsection{Impact of Different Antennas}

In order to evaluate the effect of different antennas on DeepFi performance, we consider two different versions of DeepFi: (i) DeepFi with 90 CSI values from the three antennas as input data in both phases (3-antenna DeepFi);
(ii) DeepFi with only the 30 CSI values from one of the three antennas in the training phase and estimating the position using 30 CSI values from the same antenna in the test phase (single antenna DeepFi). In addition, we set all the other parameters the same as that in the computer laboratory experiments. 

\begin{figure} [!t]
	\centering
		\includegraphics[width=3.2in]{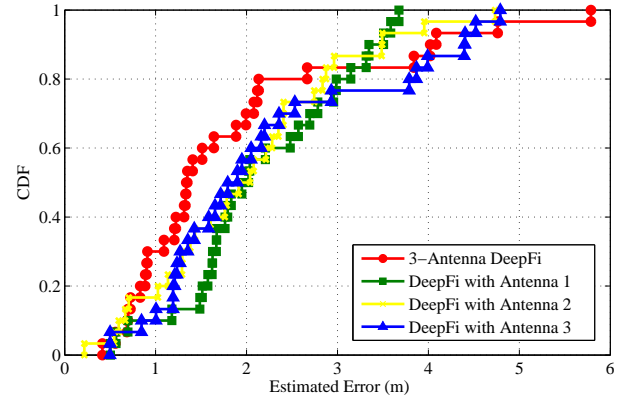}
\caption{CDF of estimated errors for DeepFi with different number of antennas.}
\label{fig:errantenna}
\end{figure}

In Fig.~\ref{fig:errantenna}, we compare these two schemes with different antennas in the training and test phases. According to the CDFs of estimation errors, 
more than 60\% of the test points in the 90-CSI scheme 
have an estimated error under 1.5 meter, while the other 30-CSI single antenna schemes 
have an estimated error under 1.5 meters for fewer than 40\% of the test points. In fact, the single antenna scheme has a mean distance error around 2.12 meters, 
while the three-antenna scheme has reduced the mean distance error to about 1.84 meters. 
Thus the 90-CSI scheme achieves better localization accuracy than the 30-CSI schemes, 
because more environment property of every sampling spot is exploited for location estimation in the test phase as the amount of CSI values is increased from 30 CSI values to 90 CSI values, thus improving the diversity of CSI samples. This experiment validates our Hypothesis 3.

Even though the 3-antennas DeepFi scheme achieves a lower mean error, it takes more time for processing the 90 CSI values as input data for each packet. We evaluate the average processing time to estimate the device position in the test phase using 100 received packets. The processing time is measured as the CPU occupation time for the Matlab program running on a laptop. In Table~\ref{tab:errantennatime}, 
we can see that the single antenna schemes take 2.3 seconds on average to estimate the device position, while the 3-antenna scheme takes around 2.5 seconds for processing the 100 packets with 90 CSI values per packet as input data to estimate the location. The difference is small, although the latter processes three times input data than that in the single antenna scheme. Although the 3-antenna DeepFi 
takes about 10 percent extra processing time, 
it can achieve a 15 percent improvement in localization precision. The latter is generally more important for indoor localization. 


\begin{table} [!t]
\centering
\caption{DeepFi Average execution time (s) versus number of antennas}
\begin{tabular}{llll}
\toprule
3-Antenna DeepFi  & Antenna 1  & Antenna 2   & Antenna 3  \\
\midrule
2.5679  &  2.3224  &  2.3043  &  2.3096 \\
\bottomrule
\end{tabular}
\label{tab:errantennatime}
\vspace{-0.1in}
\end{table}

\subsubsection{Impact of the Number of Test Packets}

In order to study the impact of the number of test packets, we design a specific experiment by utilizing different numbers of packets to evaluate their effect on both localization accuracy and execution time. In DeepFi, the laptop requests packets from the wireless router every 50 ms, i.e., at a rate of 20 packets per second. In addition, we assume that a user randomly moves with the speed of about 1 meter per second, then stays in a 1 meter square spot for 1 second, moves again, and so forth. Thus 20 packets per second are received for each test location. 

Table~\ref{tab:errpacketnum} shows the expectation and the standard deviation of localization error of 90 independent experiments. As the number of test packets is increased, the mean localization error tends to decrease. For example, the mean estimated localization error is about 1.83 meters for the case of 300 packets, which is better than the error of 1.93 meter for the case of 5 packets. This is because a large number of test packets provide a stable estimation result, thus mitigating the influence of environment noise on CSI values. Another trend is that the standard deviation of localization error will decrease as the number of packets is increased. This is because that as more samples are available, the standard deviation of samples will be decreased. On the other hand, the characteristic of clusters hidden in CSI values is revealed by increasing the number of packets, thus improving the localization accuracy.


\begin{table} [!t]
\centering
\caption{DeepFi error (m) and execution time (s) versus number of test packets}
\begin{tabular}{llllll}
\toprule
\# of Test Packets & 5 & 10 & 30 & 100 & 300 \\
\midrule
Mean Error & 1.930 & 1.915 & 1.897 & 1.898 & 1.829 \\
Std Dev    & 1.374 & 1.357 & 1.344 & 1.313 & 1.268 \\
Ave. Exe. Time &  1.744  &  1.780  &  1.949  &  2.539  &  4.205 \\
\bottomrule
\end{tabular}
\label{tab:errpacketnum}
\vspace{-0.1in}
\end{table}

In the case of using 5 test packets, although it takes less than 1/4 seconds for collecting them, DeepFi can still achieve a good performance of localization. Apart from reducing the collecting time, DeepFi using 5 test packets also simplifies the process of averaging packets in the test phase, thus significantly reducing the execution time for the online phase. We compare the average execution time of position estimation 
for 90 independent experiments based on recorded CPU occupation time for the cases of using different test packets. 
Table~\ref{tab:errpacketnum} shows that as the number of test packets is increased, the execution time also increases quickly. This is because DeepFi estimates the error of every location by averaging errors of all the test packets. For instance, the execution time with 300 packets is around 4.2 seconds, which is about 2.5 times of that with 5 packets (about 1.7 seconds). 
Therefore, even though more packets contributes to slightly improving the localization precision, we prefer to reduce the number of packets for saving collecting and processing time. 


\subsubsection{Impact of the Number of Packets per Batch}

Since deep learning utilizes $n$ packets in the test phase, how to pre-process these packets is 
important for DeepFi to reduce the computation complexity. Before the test phase in DeepFi, packets are divided into several batches, each of which contains a same number of packets. Because packets are processed in parallel in batches, we can significantly shorten the processing time when dealing with a large amount of packets. We analyze the impact of the number of packets per batch in this section. We set 1, 3, 5, 10, 50 and 100 packets per batch in the test phase with 100 collected packets. 
Again, we examine two main effects: the localization error and the test execution time. 

Table~\ref{tab:errnumcase} shows the expectation and the standard deviation of localization error with different number of packets per batch. As expected, the six experiments maintain approximately the same mean and standard deviation of errors, due to the fact that the parallel processing based on batches only averages the errors of 100 packets. 
Table~\ref{tab:errnumcase} also shows that as the number of packets per batch is increased from 1 to 10, the average execution time decreases quickly. For continuing increasing the number of packets from 10 to 100, we can find that the average computation time is approximately from 2.28 s to 2.15 s, which has smaller change. In addition, we need to average the errors over different batch data to improve the robustness of the localization results. For example, if we consider 100 packets per batch, there is only 1 batch for 100 packets, thus leading that we cannot average the errors. Thus, we employ 10 packets per batch for our DeepFi system, which not only has lower average computation time, but also higher localization results.

%

\begin{table} [!t]
\centering
\caption{DeepFi error (m) and execution time (s) versus number of packets per batch}
\begin{tabular}{lllllll}
\toprule
Pkts/Batch & 1 & 3 & 5 & 10 & 50 & 100 \\
\midrule
Mean Error & 1.839 & 1.841 & 1.846 & 1.809 & 1.852 & 1.861 \\
Std Dev     & 1.374 & 1.344 & 1.349 & 1.349 & 1.385 & 1.391 \\
Ave. Exe. &  3.453  &  2.908  &  2.568  &  2.281  & 2.220 & 2.150 \\
Time   & & & & & & \\
\bottomrule
\end{tabular}
\label{tab:errnumcase}
\vspace{-0.1in}
\end{table}

\subsection{Impact of Environment Variation}

Since the CFR changes as the indoor propagation environment varies, we examine the effect of varying propagation environment on CSI properties through two specific aspects: replaced obstacles in the room and human mobility. First, because the relative distance between the transmitter and the obstacle can affect the strength and direction of reflection of wireless signal, we consider the impact of replaced obstacles at different relative distances. In the experiment, We place a laptop and a wireless router at two fixed positions, and then add obstacles at different distances to the router, i.e., at 1 meter, 2 meters, and 3 meters locations. Then, we calculate and plot the CDF of the correlation coefficient of (i) the 90 CSI values under this cluttered environment and (ii) the 90 CSI values under the obstacle-free environment. 

\begin{figure} [!t]
	\centering
		\includegraphics[width=3.2in]{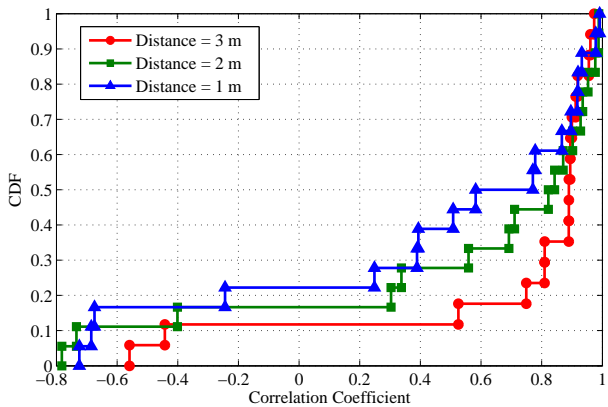}
\caption{CDF of correlation coefficient between the 90 CSI values under cluttered environment and the 90 CSI values measured without obstacles.}
\label{fig:objects}
\end{figure}

In Fig.~\ref{fig:objects}, we can see that as the distance between the obstacle and the wireless router is increased, the correlation between the two groups of 90 CSI values becomes stronger, which means that the obstacle has less impact on wireless signal transmission when it is farther away. This is due to the fact that when the obstacle is farther from the transmitter, there is lower possibility that it distorts strong signals such as the LOS signal that the laptop receives. In addition, more than 80\% of the test points have a correlation coefficient greater than 0.8 when the obstacle is 3 meters away from the wireless router. The high correlation suggests that the obstacle placed more than 3 meter has no significant impact on the 90 CSI values the laptop receives. On the other hand, when the obstacle is very close to the router, the 90 CSI values will slightly change. It leads to a smaller correlation coefficient, which affects the precision of indoor localization in the test phase based on such CSI properties. Therefore, when the obstacle arbitrarily moves in the room, its impact on CSI properties is acceptable, and high localization precision can still be achieved with DeepFi.

In addition to static obstacles, human mobility is another problem we need to consider in practical localization. The experiment of human mobility consists of two scenarios: a user randomly moves (i) near the LOS path, and (ii) near the NLOS path. To demonstrate the effect of human interference on indoor localization, we also 
plot the CDF of the correlation coefficients between (i) the 90 CSI values when a user moves near the LOS path and (ii) the 90 CSI values when a user moves near the NLOS path. 

\begin{figure} [!t]
	\centering
		\includegraphics[width=3.2in]{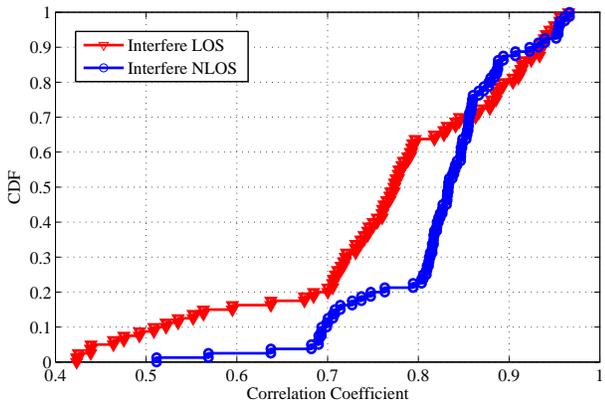}
\caption{CDF of correlation coefficients between the 90 CSI values when a user moves around the LOS path and the 90 CSI values when a user moves around the NLOS path.}
\label{fig:errhuman}
\end{figure}

We then present the human mobility experiment results
in Fig.~\ref{fig:errhuman}. It can be seen that there are only fewer than 20\% of the test points with a correlation coefficient under 0.7, if a user moves near the LOS path. On the other hand, when a user moves apart from the LOS path, approximately 20\% of the test points has a correlation coefficient under 0.8. As we can see, the correlation of the two groups of 90 CSI values if a user moves around the LOS path is weaker than that if a user moves around the reflected path, which is about 2 meter away from the wireless router. In fact, due to the stability of CSI values and high correlation coefficients for the above two scenarios, the property of the 90 CSI values will not be significantly 
affected by human mobility. Therefore, DeepFi can still achieve high localization accuracy even in a busy environment.

\subsection{Impact of the Training Grid Size} \label{subsec:sizespot}

With DeepFi, a mobile device in the test phase uses 90 CSI values it receives to search for the most similar training position. Thus, 
it is preferable that each training position possesses a unique property for the 90 CSI values. Otherwise, if most of the positions have similar CSI properties, it would be difficult to separate the matched positions from unmatched ones. As a result, these unmatched positions, which randomly scatter in the coverage space, lead to reduced localization accuracy. Therefore, in order to design a suitable training grid size for DeepFi, we study the correlation coefficient of the 90 CSI values between two neighboring training positions as the distance between them is increased. Our experiment records many pairs of positions with different distances, including 15 cm, 30 cm, 60 cm, and 120 cm. In order to mitigate the effect of the direction of the router on the correlation coefficient of the 90 CSI values,
we equally place the laptop at four directions facing north, south, west and east. 

\begin{figure} [!t]
	\centering
		\includegraphics[width=3.2in]{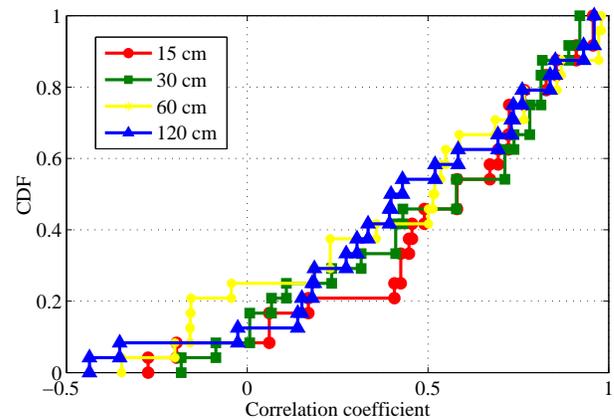}
\caption{CDF of correlation coefficient of the 90 CSI values between two adjacent training positions.}
\label{fig:gridsize}
\end{figure}

Figure~\ref{fig:gridsize} shows that as the grid size is increased, the correlation coefficient of the 90 CSI values between two neighboring positions becomes weaker. In other words, their CSI properties have less similarity due to the larger grid size. In fact, some positions even have low or negative correlation coefficients, even when the grid size is small (i.e., when they are close to each other). This is because the CFR will change as a user moves, as some multipath components may be blocked at near positions and thus some of clusters in received CSI values may be lost. If the CSI values cannot match the corresponding clusters, the correlation will obviously become low. 
%
From Fig.~\ref{fig:gridsize}, we find that the localization performance should be acceptable when the grid size is over 30 cm. 
i.e., most of the 
training positions can be separated by CSI with the 30 cm range. We thus set the grid size at about 50 cm for the training positions, so that a test position at the center of the square formed by four neighboring training positions has a distance of $50 \times \sqrt{2} / 2 = 35$ cm to the nearest training position in the worst case. A larger grid size would fail to match highly similar positions because of the scarcity of matched positions, while a smaller grid size requires redundant pre-training work.

\section{Related Work} \label{sec:RW}


There has been a considerable literature on indoor localization~\cite{LocationSurvey}. Early indoor location service systems include (i) Active Badge equipped mobiles with infrared transmitters and buildings with several infrared receivers~\cite{AB}, (ii) the Bat system that has a matrix of RF-ultrasound receivers deployed on the ceiling~\cite{Bat}, and (iii) the Cricket system that equipped buildings with combined RF/ultrasound beacons~\cite{Cricket}. All of these schemes achieve high localization accuracy due to the dedicated infrastructure. Recently, considerable efforts are made on indoor localization systems based on new hardware, with low cost, and high accuracy. These recent work mainly fall into three categories: Fingerprinting-based, Ranging-based and AOA-based, which are discussed in this chapter. 

\subsection{Fingerprinting-based Localization}

Fingerprinting-based Localization requires a training phase to survey the floor plan and a test phase to search for the most matched fingerprint for location estimation~\cite{Dynamic,Compressive}. Recently, different forms of fingerprint have been explored, including WiFi~\cite{Horus}, FM radio~\cite{FM}, RFID~\cite{RFID}, acoustic~\cite{acoustic}, GSM~\cite{GSM}, light~\cite{light} and magnetism~\cite{magnetism}, 
where WiFi-based fingerprinting is the dominant method because WiFi signal is ubiquitous in the indoor environments. 
The first work based on WiFi is RADAR~\cite{Radar}, which builds fingerprints of RSS using one or more access points. It is 
a deterministic method using KNN for position estimation. Horus~\cite{Horus} is an RSS based scheme utilizing probabilistic techniques to improve localization accuracy, where the RSS from an access point is modeled as a random variable over time and space. In addition to RSS,
channel impulse response of WiFi is considered as a location-related and stability signature, with which the fine-grained characteristics of wireless channels can be exploited to achieve higher localization accuracy. For example, FIFS~\cite{FIFS} and PinLoc~\cite{You} use CSI obtained through the off-the-shelf IWL 5300 NIC to build reliable fingerprints. 
Although these techniques achieve high localization precision, they need a large amount of calibration to build the fingerprint database via war-driving, as well as manually matching every test location 
with the corresponding fingerprint.

Crowdsourcing is proposed to reduce the burden of war-driving by sharing the load 
to multiple users. 
It consists of two main steps: (i) estimations of user trajectories, and (ii) construction of a database mapping fingerprints to user locations~\cite{crowdsourcing}. Recently, Zee~\cite{Zee} uses the inertial sensors and particle filtering to estimate a user's walking trajectory, and to collect fingerprints with WiFi data as crowd-sourced measurements for calibration. 
Similarly, LiFS~\cite{LiFS} also uses user trajectories to obtain fingerprint values and then builds the mapping between the fingerprints and the floor plan. 
Crowdsourcing can also be used to detect indoor contexts. For example, CrowdInside~\cite{CrowdInside} and Walkie-Markie~\cite{Walkie} 
can detect the shape of the floor plan and build the pathway to obtain the crowdsourced user's fingerprints. Moreover, Jigsaw~\cite{Jigsaw} and Travi-Navi ~\cite{Travi} combine vision and mobility obtained from a smartphone to build user trajectories. 
Although 
crowdsourcing does not require a large amount of calibration effort, it obtains coarse-grained fingerprints, which leads to low localization accuracy in general.

\subsection{Ranging-based Localization}

Ranging-based localization computes distances to at least three access points and leverages geometrical models for location estimation. 
These schemes are mainly classified into two categories: power-based and time-based. For power-based approaches, the prevalent log-distance path loss (LDPL) model is used to estimate distances based on RSS, where some measurements are utilized to train the parameters of the LDPL model~\cite{ZC}. For example, EZ~\cite{EZ} is a configuration-free localization scheme, 
where a genetic algorithm is used for solving the RSS-distance equations. 
The LDPL model and truncated singular value decomposition (SVD) are used to build a RSS-distance map for localization, which is adaptive to indoor environmental dynamics~\cite{ZC}. 
CSI-based ranging is proposed to overcome the instability of RSS in indoor environments. 
For instance, FILA exploits CSI from the PHY Layer to mitigate the multipath effect in the time-domain, and then trains the parameters of LDPL model to obtain the relationship between the effective CSI and distance~\cite{FILA}.

Acoustic-based ranging approaches are developed for improving indoor localization precision. H. Liu $et al$ propose a peer assisted localization technique based on smartphones to compute accurate distance estimation among peer smartphones with acoustic ranging~\cite{Push}. Centour~\cite{Centaur} leverages a Bayesian framework combining WiFi measurements and acoustic ranging, 
where two new acoustic techniques are proposed for ranging under NLOS and locating a speaker-only device based on estimating distance differences. Guoguo~\cite{Guoguo} is a smartphone-based indoor localization system, which estimates a fine-grained time-of-arrival (TOA) using beacon signals and performs NLOS identification and mitigation.

\subsection{AOA-based Localization}

Indoor localization based on angle-of-arrival (AOA) utilizes multiple antennas to estimate the incoming angles and then uses geometric relationships to obtain the user location. This technique is not only with zero start-up cost, 
but also achieves higher accuracy than other techniques such as RF fingerprinting or ranging-based systems. The challenge of this technique is how to improve the resolution of the antenna array. The recently proposed CUPID system~\cite{MP} adopts the off-the-shelf Atheros chipset with three antennas, and can obtain channel state information to estimate AOA, achieving a mean error about 20 degrees with the MUSIC algorithm. The 
relatively large error is mainly due to the low resolution of the antenna array. For high localization accuracy, the Array-Track system~\cite{Arraytrack} is implemented with two WARP systems, which are FPGA-based software defined radios. It incorporates a rectangular array of 16 antennas to compute the AOA, and then uses spatial smoothing to suppress the effect of multipath on AOA. However, Array-Track requires a large number of antennas, 
which is 
generally not available for commodity mobile devices. 

On the other hand, some systems, such as LTEye~\cite{LTE}, Ubicarse~\cite{Zero}, Wi-Vi~\cite{Dude}, and PinIt~\cite{Wall}, use Synthetic Aperture Radar (SAR) to mimic an antenna array to improve the resolution of angles. 
the main idea of SAR is to use a moving antenna to obtain signal snapshots as it moves along a trajectory, and then to utilize these snapshots to mimic a large antenna array along the trajectory. However, it requires accurate control of the speed and trajectory by using a moving antenna placed on an iRobot Create robot. 


\section{Conclusion}\label{sec:conC}

In this paper, we presented DeepFi, a deep learning based indoor fingerprinting scheme that uses CSI information. In DeepFi, CSI information for all the subcarriers and all the antennas are collected through the device driver and analyzed with a deep network with four hidden layers. 
Based on the three hypotheses on CSI, we proposed to use the weights in the deep network to represent fingerprints, and incorporated a greedy learning algorithm for weight training 
to reduce complexity. In addition, a probabilistic data fusion method based on the RBF was developed for online location estimation. 
The proposed DeepFi scheme was validated in two representative indoor environments, and was found to outperform several existing RSS and CSI based schemes in both experiments. 
%
We also examined the effect of different parameters 
and varying propagation environments on DeepFi performance, 
and found that DeepFi can achieve good performance under such scenarios.

\bibliographystyle{IEEEtran}
\bibliography{handoff_femto}


\begin{biography}
[{\includegraphics[width=1in,height=1.25in,clip,keepaspectratio]{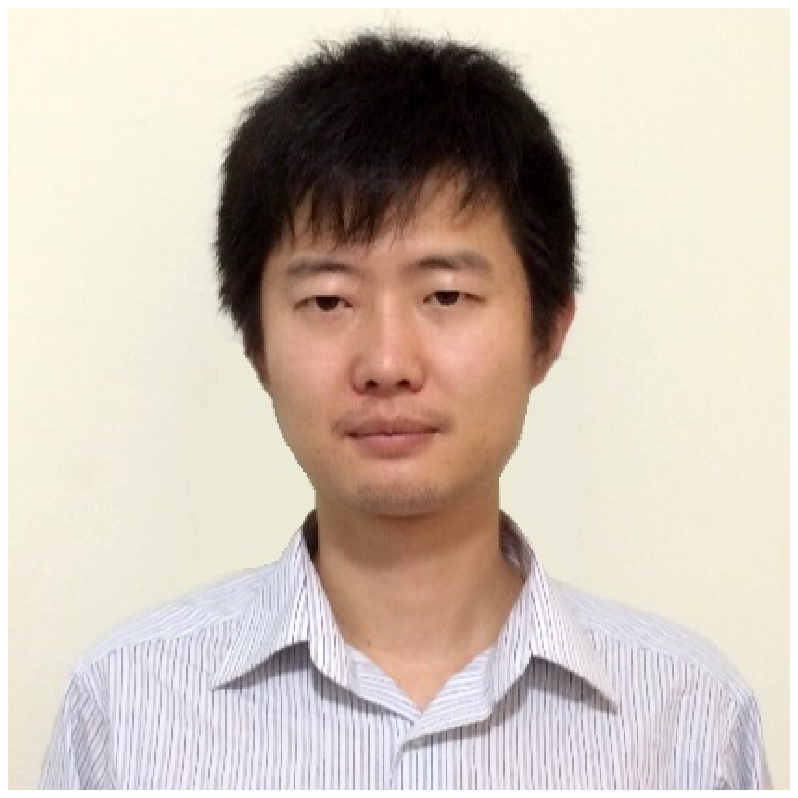}}]{Xuyu Wang} (S'13) received the M.S. in Signal and Information Processing in 2012 and the B.S. in Electronic Information Engineering in 2009, both from Xidian University, Xi'an, China. Since 2013, he has been pursuing a Ph.D. degree in the Department of Electrical and Computer Engineering, Auburn University, Auburn, AL, USA. His research interests include indoor localization, deep learning, wireless communications, software defined radio, and big data. He is a recipient of a Woltolsz Fellowship at Auburn University, and a co-recipient of the Second Prize of Natural Scientific Award of Ministry of Education, China in 2013. 
\end{biography}

\vfill

\begin{biography}
[{\includegraphics[width=1in,height=1.25in,clip,keepaspectratio]{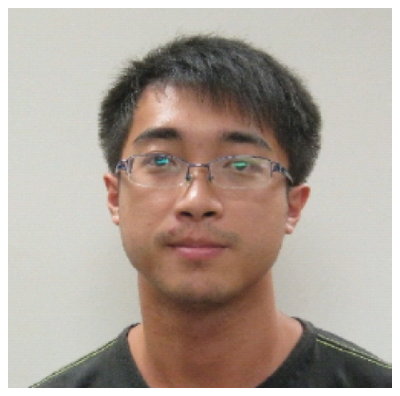}}]{Lingjun Gao} (S'14) received his M.S. in Electrical and Computer Engineering from Auburn University, Auburn, AL, USA in 2015, and his B.E. in Electrical Engineering from Civil Aviation University of China, Tianjing, China in 2013. Currently, he is a Data Engineer with DataYes, Inc. in Shanghai, China. His research interests include machine learning, indoor localization, and testbed implementation.  
\end{biography}

\vfill

\begin{biography}
[{\includegraphics[width=1in,height=1.25in,clip,keepaspectratio]{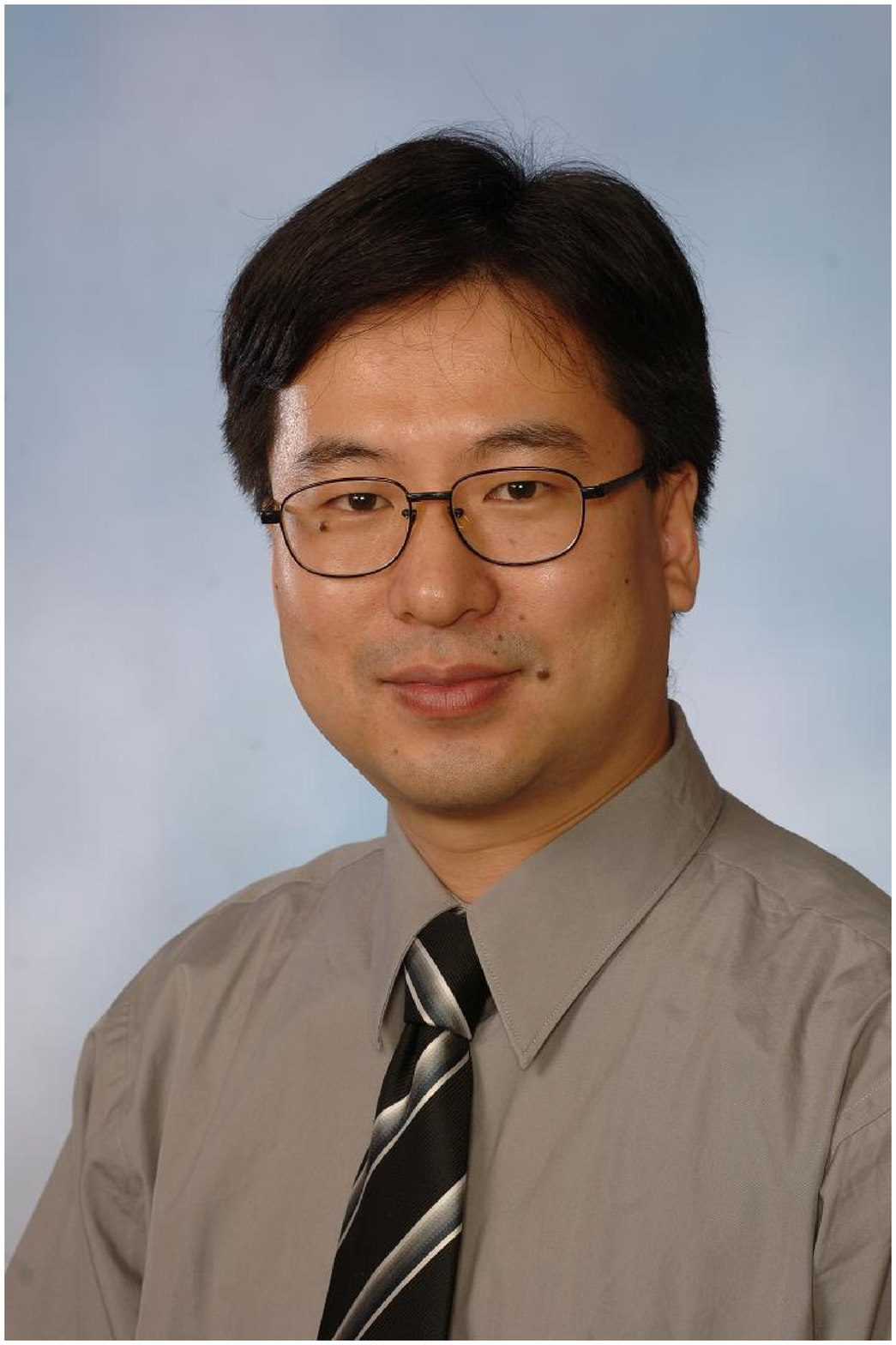}}]{Shiwen Mao} (S'99-M'04-SM'09) received his Ph.D. in electrical and computer engineering from Polytechnic University, Brooklyn, NY in 2004. Currently, he is the Samuel Ginn Distinguished Professor and Director of Wireless Engineering Research and Education Center (WEREC) 
at Auburn University, Auburn, AL, USA. His research interests include wireless networks and multimedia communications. 
He is a Distinguished Lecturer of IEEE Vehicular Technology Society. He is on the Editorial Board of IEEE Transactions on Multimedia, IEEE Internet of Things Journal, and IEEE Communications Surveys and Tutorials, and IEEE Multimedia, among others. He serves as Area TPC Chair of IEEE INFOCOM 2017 and 2016, Technical Program Vice Chair for Information Systems (EDAS) of IEEE INFOCOM 2015, symposium/track co-chair for many conferences, including IEEE ICC, IEEE GLOBECOM, ICCCN, 
among others, Steering Committee Voting Member for IEEE ICME and AdhocNets, and in various roles in the organizing committees of many conferences. He is the Vice Chair--Letters \& Member Communications of IEEE ComSoc Multimedia Communications Technical Committee. He received the 2015 IEEE ComSoc TC-CSR Distinguished Service Award, the 2013 IEEE ComSoc MMTC Outstanding Leadership Award, and the NSF CAREER Award in 2010. He is a co-recipient of the IEEE GLOBECOM 2015 Best Paper Award, the IEEE WCNC 2015 Best Paper Award, the IEEE ICC 2013 Best Paper Award, and the 2004 IEEE Communications Society Leonard G. Abraham Prize in the Field of Communications Systems.
\end{biography}

\vfill

\begin{biography}
[{\includegraphics[width=1in,height=1.25in,clip,keepaspectratio]{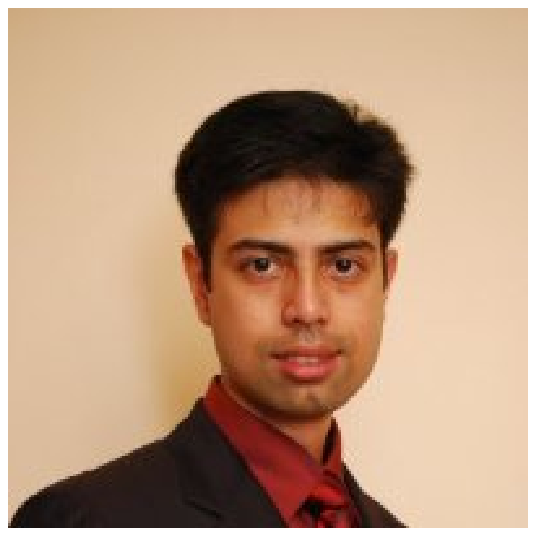}}]{Santosh Pandey} received his B.S. in electrical engineering from the University of Mumbai in 2002, and his M.S. and Ph.D. from Auburn University in 2007. He joined Cisco Systems in 2007 as a system engineer, where he has primarily worked on location algorithms in wireless networks. He has additionally worked on handoff and handover algorithms and simulations for fixed mobile convergence. He actively participates in IEEE 802.11 Task Group AE, which aims to prioritize management frames in 802.11 networks.
\end{biography}

\vfill

\end{document}